\documentclass[aps,twocolumn,superscriptaddress]{revtex4}

\usepackage[paperwidth=210mm,paperheight=297mm,centering,hmargin=2cm,vmargin=2.5cm]{geometry}

\usepackage{amsmath,amssymb}
\usepackage[pdftex]{graphicx}
\usepackage{array}

\usepackage{xcolor}

\definecolor{bluemoi}{rgb}{0.25,0.50 ,0.75} 

\usepackage{hyperref}
\hypersetup{
bookmarksopen=false,
pdftitle=" ",
pdfauthor=" ", 
pdfsubject=" ", 
pdftoolbar=true, 
pdfmenubar=true, 
pdfhighlight=/O, 
colorlinks=true, 
pdfpagelayout=SinglePage, 
pdffitwindow=true, 
linkcolor=bluemoi, 
citecolor=bluemoi, 
urlcolor=bluemoi 
}

\makeatletter
\renewcommand{\figurename}{Figure}
\makeatother
\makeatletter
\renewcommand{\fnum@figure}{\small\textbf{\figurename~\thefigure}}
\makeatother

\begin{document}

\title{Human diffusion and city influence}

\author{Maxime Lenormand\email{maxime@ifisc.uib-csic.es}}\affiliation{Instituto de F\'{\i}sica Interdisciplinar y Sistemas Complejos IFISC (CSIC-UIB), 07122 Palma de Mallorca, Spain}
\author{Bruno Gon{\c c}alves}\affiliation{Aix Marseille Universit{\' e}, Universit{\'e } de Toulon, CNRS, CPT, UMR 7332, 13288 Marseille, France}
\author{Ant{\`o}nia Tugores}\affiliation{Instituto de F\'{\i}sica Interdisciplinar y Sistemas Complejos IFISC (CSIC-UIB), 07122 Palma de Mallorca, Spain}
\author{Jos{\'e} J. Ramasco}\affiliation{Instituto de F\'{\i}sica Interdisciplinar y Sistemas Complejos IFISC (CSIC-UIB), 07122 Palma de Mallorca, Spain}

\begin{abstract} 
Cities are characterized by concentrating population, economic activity and services. However, not all cities are equal and a natural hierarchy at local, regional or global scales spontaneously emerges.  In this work, we introduce a method to quantify city influence using geolocated tweets to characterize human mobility.  Rome and Paris appear consistently as the cities attracting most diverse visitors. The ratio between locals and non-local visitors turns out to be  fundamental for a city to truly be global. Focusing only on urban residents' mobility flows, a city to city network can be constructed. This network allows us to analyze centrality measures at different scales. New York and London play a predominant role at the global scale, while urban rankings suffer substantial changes if the focus is set at a regional level.  
\end{abstract}

\maketitle

Ever since Christaller proposed the central place theory in the 30's \cite{christaller33}, researchers have worked to understand the relations and competition between cities leading to the emergence of a hierarchy. Christaller envisioned an exclusive area surrounding each city at a regional scale to which it provided services such as markets, hospitals, schools, universities, etc. The services display different level of specialization, inducing thus a hierarchy among urban areas according to the type of services offered. In addition, this idea naturally brings an equidistant distribution of urban centers of similar category as long as no geographical constraints prevents it. Still, in the present globalized world relations between cities go much beyond mere geographical distance. In order to take into account this fact, it was necessary to introduce the concept of world city \cite{friedmann82}. These are cities that concentrate economic warehouses like the headquarters of large multinational companies or global financial districts, of knowledge and innovation as the cutting edge technological firms or universities, or political decision centers, and that play an eminent role of dominance over smaller, more local, counterparts. The concept of global city is, nevertheless, vague and in need of further mathematical formalization. This is attained by means of so-called world city networks, in which each pair of cities is linked whether they share a common resource or interchange goods or people \cite{Berry64,taylor95,rimmer98,Pumain2000,taylor01}. For instance, a link can be established if two cities share headquarters of the same company \cite{taylor01,derudder03,derudder04}, if both are part of good production chains \cite{brown10}, interchange finance services \cite{bassens10},  internet data \cite{neal10} or if direct flights or boats connect them \cite{taylor95,zook05,derudder05,derudder08}. Centrality measures are then applied to the network and a ranking of the cities naturally emerges. Due in great part to their geographical locations and traditional roles as trans-Atlantic bridges, New York and London are typically the top rankers in many of these studies \cite{rimmer98,derudder04,derudder05}. There are, however, inconsistencies in terms of the meaning and stability  of the results obtained from different networks or with different centrality measures \cite{derudder05,allen10} and a more organic and stable definition is needed.    

Here we use information and communication technologies (ICT) to approach the problem from a different perspective. How long would information originating from a given city require to reach any other city if were to pass from person to person only through face to face conversations? Or, in other words, what is the likelihood that that information reaches a certain distance away after a given time period. In this thought experiment, the most {\em central} place in the world would simply be the one where the message can reach everywhere else in the shortest amount of time. This view allows us to easily define a temporal network of influence.

We perform this analysis by empirically observing how people travel worldwide and using that as a proxy for how quickly our message would be able to spread. The recent popularization and affordability of geolocated ICT services and devices such as mobile phones, credit or transport cards gets registered generating a large quantity of real time data on how people move \cite{brockmann06,gonzalez08,balcan09,noulas12,bagrow12,grabowicz13,lenormand14, tizzoni14, jurdak14}. This information has been used to study questions such as interactions in social networks \cite{java07,krishnamurthy08,huberman08,grabowicz12}, information propagation \cite{ferrara13}, city structure and land use \cite{reades07,reades09,Cheng2010,soto11,frias12,noulas13,louail14,grauwin14,lenormand14,Franca2014,louail15}, or even road and long range train traffic \cite{lenormand14b}. It is bringing a new era in the so-called Science of Cities by providing a ground for a systematic comparison of the structure of urban areas of different sizes or in different countries \cite{bettencourt07,batty08,bettencourt10, batty13,bettencourt13,Adnan2014,louail14,grauwin14,louail15}.  Data coming from credit cards and mobile phones are usually constrained to a limited geographical area such as a city or a country, while those coming from online social media as Twitter, Flickr or Foursquare can refer to the whole globe. This is the reason why we focus here on geolocated tweets, which have already proven to be an useful tool to analyze mobility between countries \cite{hawelka13} and provide the ideal framework for our analysis. 

\begin{figure}
\begin{center}
\includegraphics[width=8.6cm]{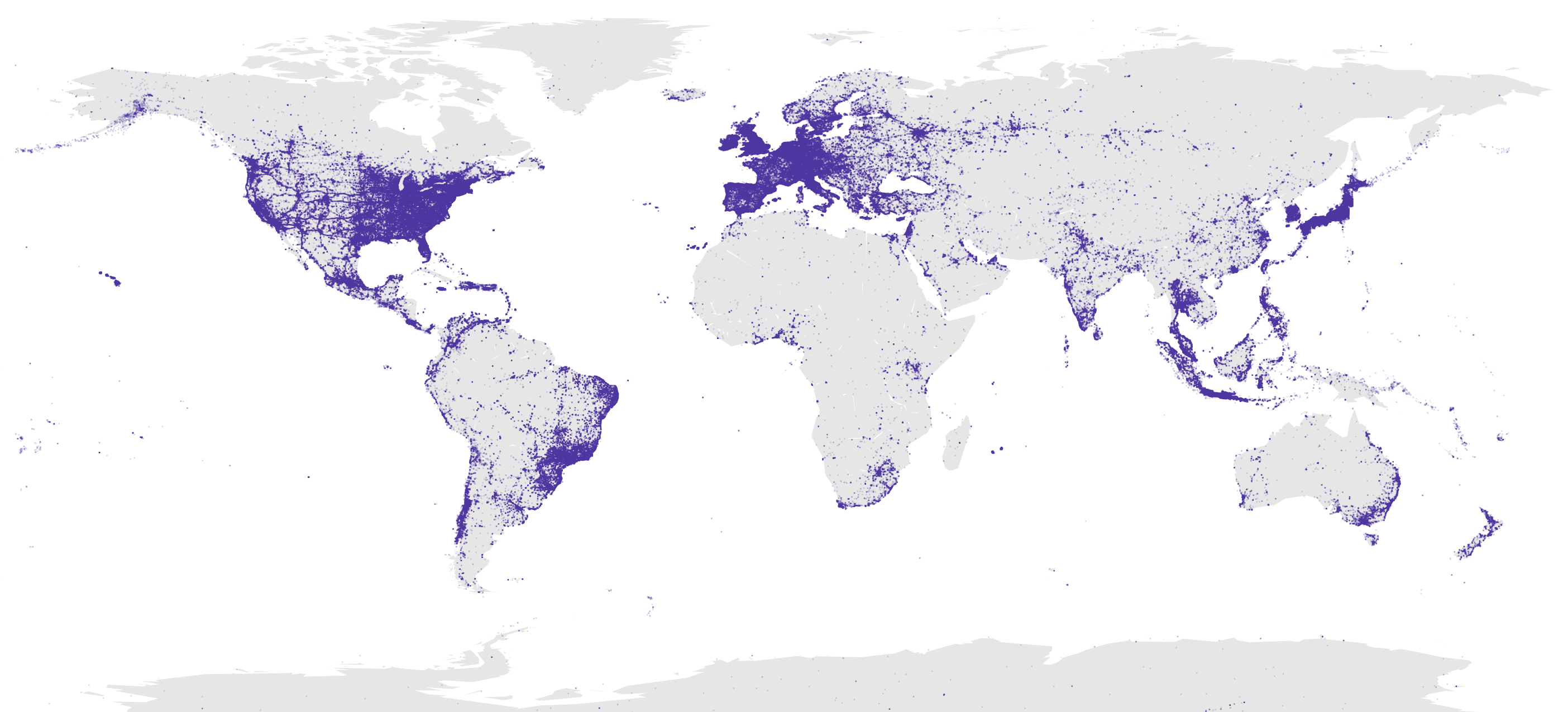}
\caption{\textbf{Positions of the geolocated tweets.} Each tweet is represented as a point on the map location from which it was posted.\label{map}}
\end{center}
\end{figure}

In particular, we select $58$ out of the most populated cities of the world and analyze their influence in terms of the average radius traveled and the area covered by Twitter users visiting each of them as a function of time. Differences in the mobility for local residents and external visitors are taken into account, in such a way that cities can be ranked according to the extension covered by the {\em diffusion} of visitors and residents, taken both together and separately, and by the attractiveness they exhibit towards visitors. Finally, we also consider the interaction between cities, forming a network that provide a framework to study urban communities and the role cities play within their own community (regional) versus a global perspective.

\section*{MATERIALS AND METHODS}

\subsection*{Twitter Dataset}

Our database contains $21,017,892$ tweets geolocated worldwide written by  $571,893$ users in the temporal period ranging from October $2010$ to June $2013$ ($1000$ days). There are on average $36$ tweets per user. Non-human behaviors or collective accounts have been excluded from the data by filtering out users traveling faster than a plane ($750$ km/h). For this, we have computed the distance and the time spent between two successive geolocated tweets posted by the same user. The geographical distribution of tweets is plotted in Figure \ref{map}. The distribution matches population density in many countries, although it is important to note that some areas are under-represented as, for example, most of Africa and China. 

\begin{figure}
\begin{center}
\includegraphics[width=8.6cm]{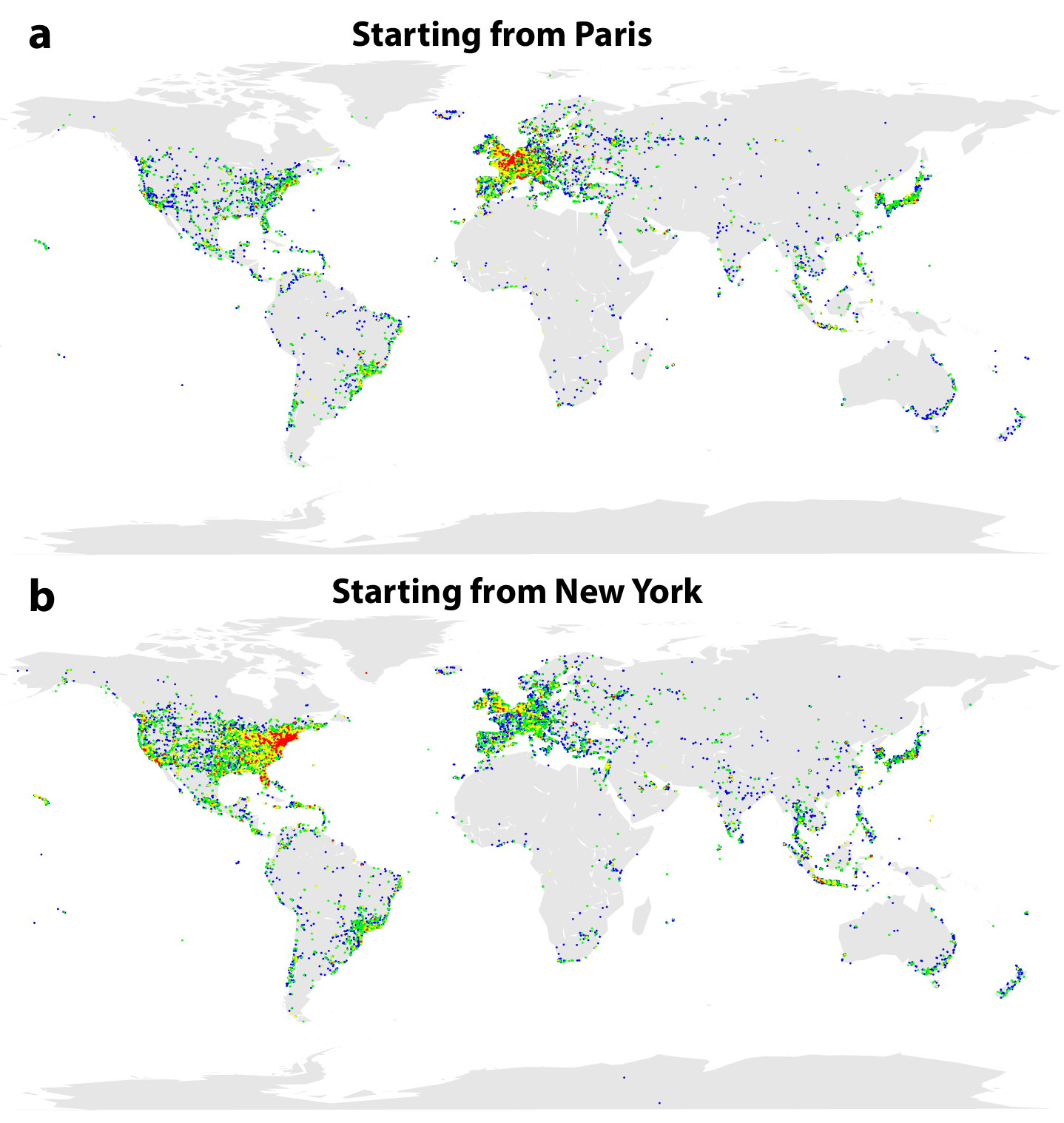}
\caption{\textbf{Geolocated tweets of users who have been at least once in Paris (a) and New York (b).} The color changes according to the number of days $\Delta_t$ since the first passage in the city. In red, one day; In yellow, between 1 and 10 days; In green, between 10 and 100 days; And in blue, more than 100 days. \label{Diff}}
\end{center}		
\end{figure}  

We take as reference $58$ cities around the world (see Table S1 in Appendix for a detailed account) that are both highly populated (most are among the $100$ most populated cities in the world) and have a sufficiently large number of geolocated Twitter users. To avoid distortions imposed by different spatial scales and urban area definitions that can be problematic \cite{Erneson2014,arcaute14}, we operationally defined each city to be a circle of radius $50$ km around the respective City Hall.   
  
In order to assess the influence of a city, we need to characterize how users travel after visiting it. To do so, we consider the tweets posted by user $\upsilon$ $\Delta_t$ days after visiting city $c$. In Figure \ref{Diff}, the locations of geolocated tweets are plotted according to the number of days since the first visit in Paris and New York as an example. Not surprisingly, a large part of the tweets are concentrated around these cities but one can observe how users eventually diffuse worldwide. 

\subsection*{Definition of the user's place of residence}

To identify the Twitter users' place of residence, we start by discretizing the space. To do so, we divide the world using a grid composed of $100 \times 100$ square kilometers cell in a cylindrical equal-area projection. In total there are approximately $5,000$ inhabited cells in our dataset. The place of residence of a user is a priori given by the cell from which he or she has posted most of his/her tweets. However, to avoid selecting users who did not show enough regularity, we consider only those users who posted at least one third of their tweets form the place of residence (representing more than 95\% of the overall users). For each city, the number of valid users as well as the number of tweets posted from their first passage in the city are provided in Table S1 in Appendix.

We can now determine for each city if a user is resident (local user) or a visitor (non-local user). To do so, we compute the average position of the tweets posted from his/her cell of residence. If this position falls within the city boundaries (circle of radius $50$ km around the City Hall) the user is considered as a local and as a non-local user otherwise.

\subsection*{Metrics to assess city influence}

\begin{figure}
\begin{center}
\includegraphics[width=8.6cm]{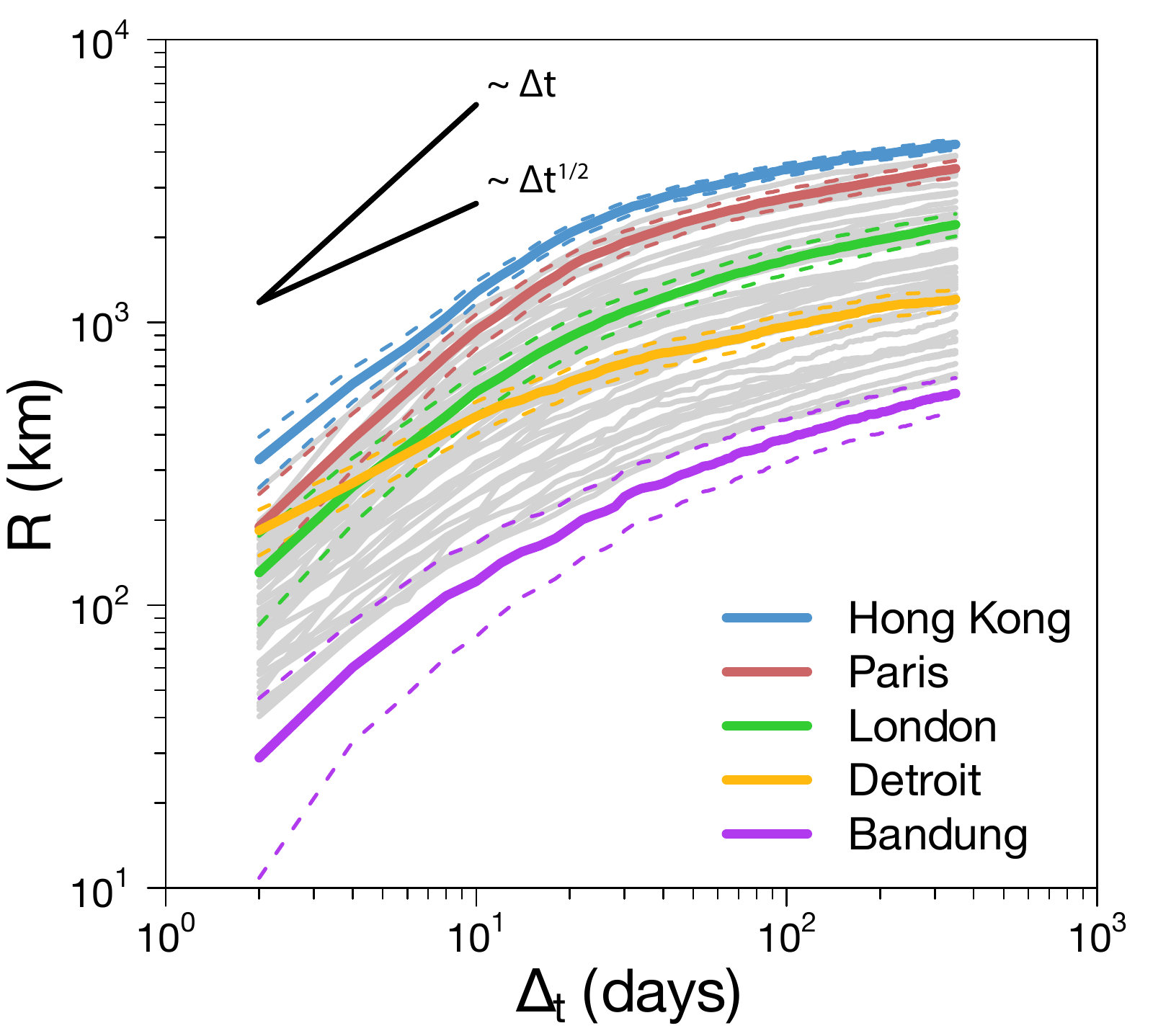}
\caption{\textbf{Evolution of the average radius.} Each curve represents the evolution of the average radius $R$ averaged over $100$ independent extractions of a set of $u = 300$ users as a function of the number of days $\Delta_t$ since the first passage in the city. In order to show the general trend, each gray curve corresponds to a city. The evolution of the radius for several cities is highlighted, such as the top and bottom rankers or representatives of the two main detected behaviors. Curves with a linear and square root growth are also shown as a guide to the eye. The dashed lines represent the standard deviation. \label{Dist}}
\end{center}
\end{figure}

We select a fixed number of users $u$ in each city at random and track their displacements in a given period of time $\Delta_t$ since their first tweet from it. Since the results might depend on the specific set of users chosen, we average over $100$ independent user extractions. As shown in Figure S2 in Appendix, the longer $\Delta_t$ is, the lower is the population of users who remain active, so we must establish a tradeoff between number of users and activity time. Unless otherwise stated we set $u = 300$ and $\Delta_t = 350$ days in the discussion that follows.  

\subsubsection*{Average radius}

There are different aspects to take into account when trying to define how to properly measure the influence of a city due to Human Mobility. We start our discussion by considering the average radius traveled by Twitter users since their first tweet from a city $c$. We tracked for each user the positions from which he or she tweeted after visiting $c$, and compute the average distance from these locations to the center of $c$. The average radius, $R$, is then defined as the average over all the $u$ users of their individual radii. 

The average radius is informative but can be biased by the geography. Cities that are in relatively isolated positions such as islands may have a high average radius just because a long trip is the only option to travel to them. To avoid this effect, we define the normalized average radius $\tilde{R}$ of a city $c$ as the ratio between $R\left(c\right)$ and the average distance of all the Twitter users' places of residence to $c$ (Figure S4 in Appendix). 

\begin{figure*}
\begin{center}
\includegraphics[width=\linewidth]{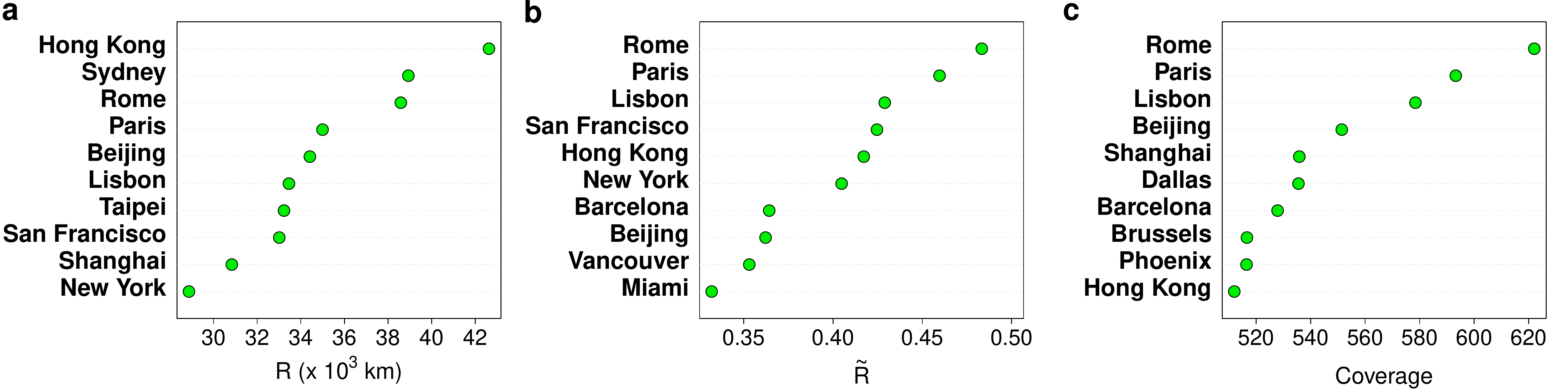}
\caption{\textbf{Rankings of the cities according to the average radius and the coverage.} (a) Top $10$ cities ranked by the average radius $R$. (b) Top $10$ cities ranked by the normalized average radius $\tilde{R}$. (c) Top $10$ cities ranked by the coverage (number of visited cells). All the metrics are averaged over $100$ independent extractions of a set of $u = 300$ users. \label{Rank}}
\end{center}			
\end{figure*}

\subsubsection*{Coverage}

One possible way to overcome the limitation of the average ratio defined above is to discard geographic coherence all together and simply measure the geographical area covered by those users, regardless of the distance at which it might be located from the originating city. In order to estimate the area cover by the users, the world surface has been divided in cells of $100 \times 100$ square kilometers as we have done to identify the users' place of residence. By tracking the movements of the set of users passing through each city, we count the number of cells from which at least a tweet has been posted and define coverage as this number. This metric has the clear advantage of not being sensitive to isolated locations but it still does not consider how specific cells, specially the ones corresponding to other important cities, are visited much more often than others.

\section*{RESULTS}

\subsection*{Comparing the influence of cities}

We start by taking the perspective from the city to the world and compare how effective the cities are as starting points for the Twitter users' diffusion. The evolution of the average radius as a function of the time is plotted in Figure \ref{Dist} for the $58$ cities. The curves of the log-log plot show an initial fast increase followed by a much slower growth after approximately $15-20$ days. The presence of these two regimes is mainly due to the presence of non-local users as it can be observed in Figure S5 in Appendix. In the initial phase, the radius grows for all the cities at a rhythm faster than the square root of time, which is the classical prediction for $2D$ Wiener diffusion \cite{weiss1994}. This is not fully surprising since the users' mobility is better described by Levy flights than by a Wiener process. Still the differences between cities are remarkable. There are two main behaviors: the radius for cities such as Detroit grows slowly, while others like Paris show an increase that is close to linear. After this initial transient, the average radius enters in a regime of slow growth for all the cities that is even slower than $\sqrt{\Delta_t}$. This implies that the long displacements by the users are concentrated in the first month, period during which the non-local users come back home, after which the exploration becomes more localized. Even though the curves of different cities may cross in the first regime, they reach a relatively stable configuration in the second one. We can see that the top ranker in terms of capacity of diffusion is Hong Kong for the whole time window considered and the bottom one is Bandung (West Java, Indonesia). 

The top $10$ cities according to the average radius are plotted in Figure \ref{Rank}a. It is worth noting New York only appears in the last position, in contrast to previously published rankings based on different data \cite{rimmer98,derudder04,derudder05}. Many cities on the top are in the Pacific Basin (Hong-Kong, Sydney, Beijing, Taipei, San Francisco and Shanghai), which is clear evidence for the impact of geography on $R$. We take geographical effects into account by calculating the normalized radius $\tilde{R}$ as shown in Figure \ref{Rank}b. With this correction, the top cities are Rome, Paris and Lisbon. These cities are located in densely populated Europe but still manage to send travelers further away than any other, a proof for their aptitude as sources for the information spreading thought experiment described in the introduction. Actually, all cities in the Top $10$ set are also able to attract visitors at a worldwide scale, some are relatively far from other global cities and/or they may be the gate to extensive hinterlands (China). The same ranking for the coverage is shown in Figure \ref{Rank}c. Even though these two metrics are strongly correlated (see Figure S6 in Appendix) there are still some significant differences indicating that they are able to capture different information. The top cities, however, are again Rome, Paris and Lisbon probably due to a combination of the factors explained above. It should also be noted that even though the users extraction is stochastic and the rankings can variate slightly from a realization to another (see Figure S7 in Appendix), the ranking is stable when performed on the average over several realizations it becomes stable (Figure S8 in Appendix).      

\subsection*{Local versus non-local Twitter users}

\begin{figure}
\begin{center}
\includegraphics[width=\linewidth]{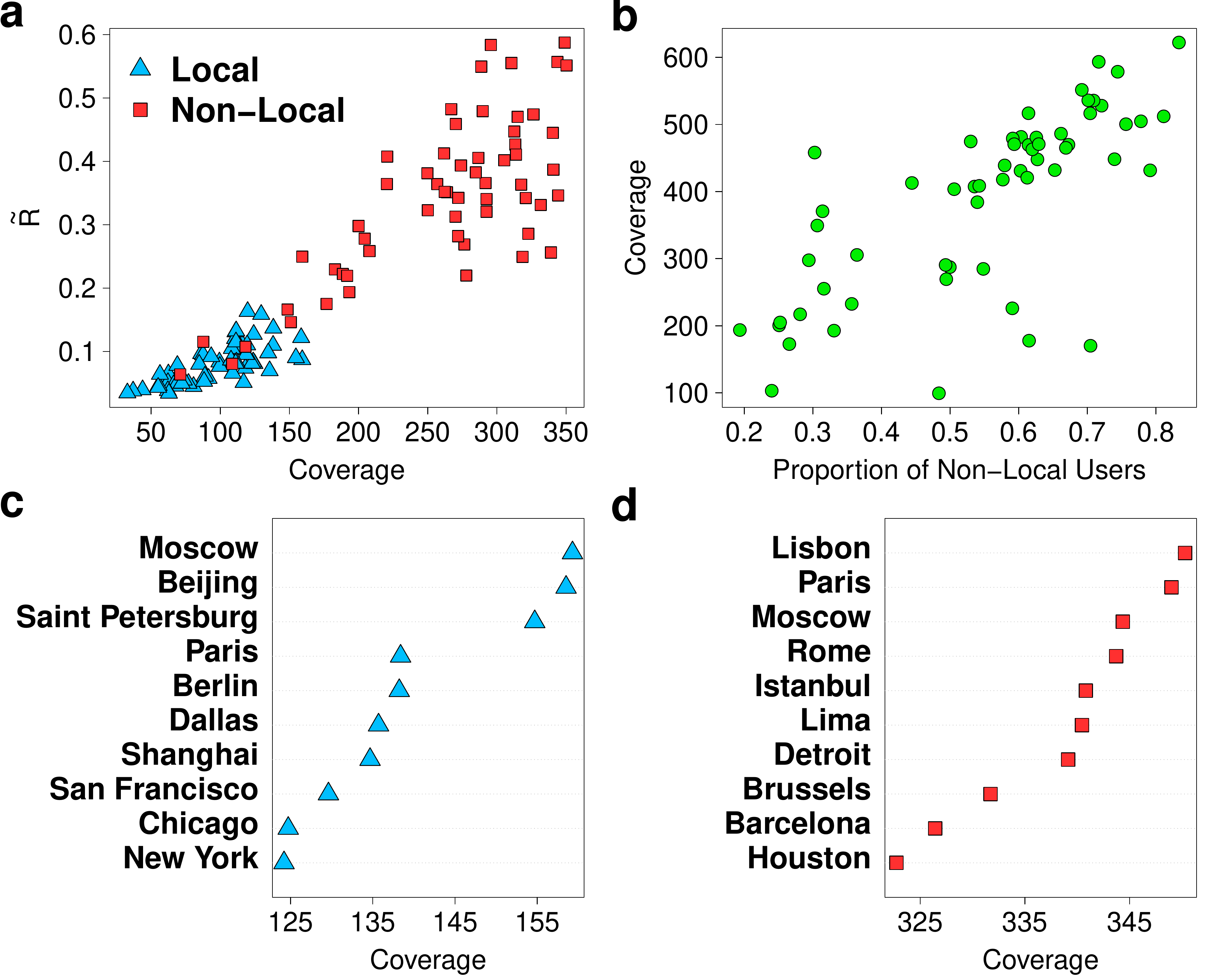}
\caption{\textbf{Relation between local and non-local users.} (a) Scatter-plot of $\tilde{R}$ as a function of the coverage for locals (blue triangles) and non-locals (red squares).  (b) Coverage as a function of the proportion of non-local Twitter users. (c) Top 10 ranking cities based only on local users according to the coverage. (d) The same ranking but based only on the movements of non-local users. In all the cases, the number of local and non-local users extracted is $u = 100$ for every city and all the metrics are averaged over $100$ independent extractions. \label{locvsnoloc}}
\end{center}	
\end{figure}

We have yet to take into account that individual residing in a city might behave differently from visitors. We consider a user to be a resident of a city if most of his/her tweets are posted from it. Otherwise, he/she is seen as an external visitor. Residents of the $58$ cities we consider have a significantly lower coverage (about $96$) than visitors (about $260$). This means that the locals move toward more concentrated locations, such as places of work or the residences of family and friends, while visitors have a comparatively higher diversity of origins and destinations.

The difference between locals and non-locals is even more dramatic when the normalized radius, $\tilde{R}$, for each city is plotted as a function of the coverage for both types of users in Figure \ref{locvsnoloc}a. Two clusters clearly emerge, showing that the locals tend to move less than the visitors. Such difference between users is likely to be behind the change of behavior in the temporal evolution of the average radius detected in Figure \ref{Dist}, and introduces the ratio of visitors over local users as a relevant parameter to describe the mobility from a city. Indeed, visitors contribute the most for the radius and the area covered (see  Figure \ref{locvsnoloc}b for the coverage) while residents contribute most to the local relevance of a city (Figure \ref{locvsnoloc}c for the coverage and Figure S10a in Appendix for $\tilde{R}$).  The top rankers in this classification are Hong Kong and San Francisco in $\tilde{R}$ and Moscow and Beijing in the coverage. All of them are cities that may act as gates for quite extense hinterlands. The rankings based on non-locals (Figures \ref{locvsnoloc}d for the coverage and Figure S10b in Appendix for $\tilde{R}$) get us back the more common top rankers such as Paris, New York and Lisbon.

\subsection*{City attractiveness}

Thus far, we have considered a city as origin and analyzed how people visiting it diffuse across the planet. We now consider the attractiveness of a city by taking the opposite point of view and analyzing the origins of each user seen within the confines of a city. We modify the two metrics defined above to consider the normalized average distance of the users' residences (represented by the centroid of the cell of residence) to the center of the considered city $c$ and the number of different cells where these users come from. In this case, the to metrics are averaged over $100$ independent extractions of $u=1000$ Twitter users. The resulting rankings depict the attractiveness of each city from the perspective of external visitors: How far are people willing to travel to visit this city? The Top $10$ cities are shown in Figure \ref{Attractivity} for the coverage (see Figure S11 in Appendix for the normalized average radius). Rome, Paris and Lisbon are also quite consistently the top rankers in terms of attractiveness to external visitors.

\begin{figure}
\begin{center}
\includegraphics[width=\linewidth]{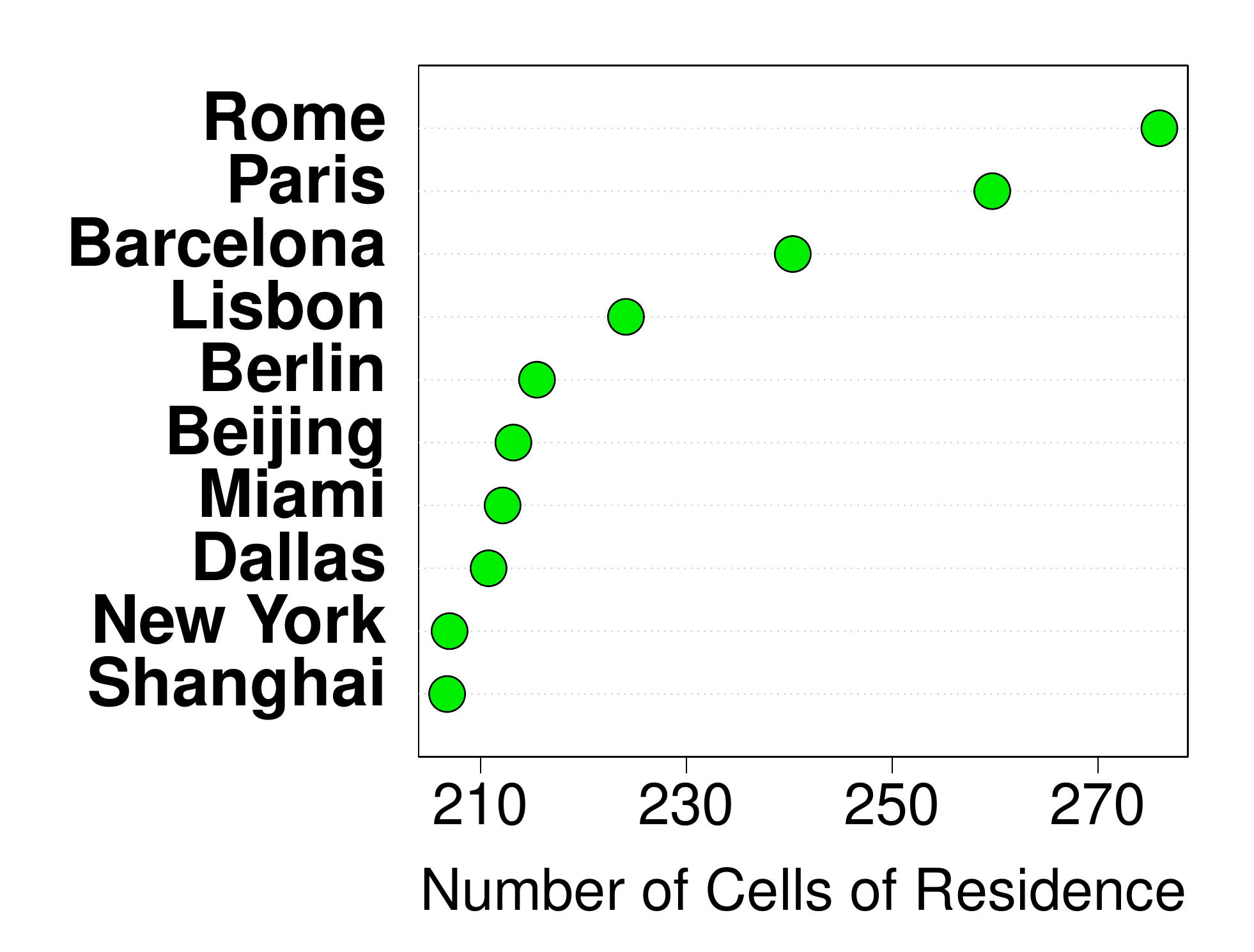} 
\caption{\textbf{City attractiveness.} Top 10 cities ranked by the number of distinct cells of residence for $u=1000$ Twitter users drawn at random. The metric is averaged over $100$ independent extractions. \label{Attractivity}}
\end{center}
\end{figure}

\subsection*{A network of cities}

\begin{figure*}
 \begin{center}
\includegraphics[width=\textwidth]{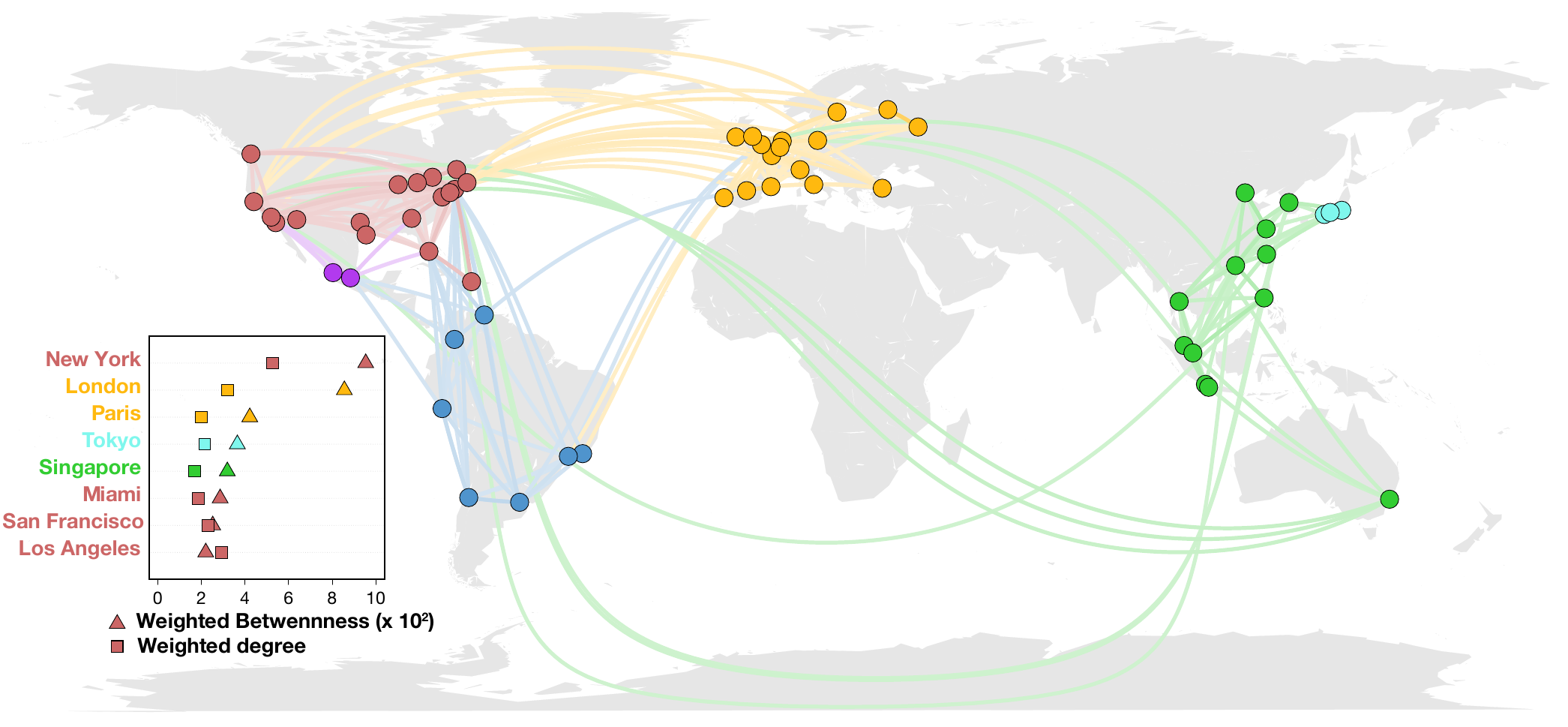}
\caption{\textbf{Mobility network.} Local Twitter users mobility network between the $58$ cities. Only the flows representing the top $95\%$ of the total flow have been plotted. The flows are drawn from the least to the greatest. The inset shows the top $8$ cities ranked by weighted betweenness and weighted degree. \label{Net}}
\end{center}
\end{figure*}

Finally, we complete our analysis by considering travel between the $58$ selected cities. We build a network connecting the $58$ cities under consideration where the directed edge from city $i$ to city $j$ has a weight given by the fraction of local Twitter users in the city $i$ which were observed at least once in city $j$. For simplicity, in what follows, we consider only local users who left their city at least once. This network captures the strength of connections between cities allowing us to analyze the communities that naturally arise due to human mobility. Using the OSLOM clustering detection algorithm \cite{lancichinetti10,lancichinetti11} we find $6$ communities as shown in Figure \ref{Net}. These communities follow approximately the natural boundaries between continents: two communities in North and Center America, one community in South America, another in Europe, two communities in Asia (Japan and rest of Asia plus Sydney), indicating that they correspond to economic, cultural and geographical proximities. Similar results were obtained using the Infomap \cite{rosvall08} cluster detection algorithm, confirming the robustness of the communities detected. 

\begin{table}[!ht]
\begin{center}
\begin{tabular}{l|l}
\hline
\multicolumn{2}{c}{\textbf{North America}}\\
\hline
Global Ranking & Regional Ranking\\
\hline
1. New York (1)	&	1. New York	\\
2. Miami (6)	&	2. Los Angeles	\\
3. San Francisco (8)	&	3. Chicago	\\
4. Los Angeles (9)	&	4. Toronto	\\
5. Chicago (18)	&	5. Detroit	\\
6. Toronto (19)	&	6. Miami	\\
7. San Diego (23)	&	7. Dallas	\\
8. Detroit (25)	&	8. San Francisco	\\
9. Montreal (26)	&	9. Washington	\\
10. Atlanta (27)	&	10. Atlanta	\\
\hline
\multicolumn{2}{c}{$\,$}\\
\hline
\multicolumn{2}{c}{\textbf{Europe}}\\
\hline
Global Ranking & Regional Ranking\\
\hline
1. London (2)	&	1. London	\\
2. Paris (3)	&	2. Paris	\\
3. Madrid (10) &	3. Moscow	\\
4. Barcelona (11)	&	4. Barcelona	\\
5. Moscow (16) &	5. Berlin	\\
6. Berlin (20) &	6. Rome	\\
7. Rome (21) &	7. Madrid	\\
8. Amsterdam (24) &	8. Lisbon	\\
 9. Lisbon (38) &	9. Amsterdam	\\
10. Milan (40) &	10. Saint Petersburg	\\
\hline
\end{tabular}
\caption{\textbf{Comparison of the regional and the global betweenness rankings.} In parenthesis the total global ranking position of each city.\label{Betweenness}}		
\end{center}
\end{table}

With these empirical communities in hand we can now place each city into a local as well as a global context. In a network context, the importance of each node can be measured in different ways. Two classical measures are the strength of a node \cite{barrat04} and the weighted betweenness \cite{newman01,brandes01}. Given the way we defined our network above, these correspond, roughly, to the fraction of local users that travel out of a city and how important that city is in connecting travelers coming from other cities to their final destinations. In the inset of the Figure \ref{Net}, we analyze the ranking resulting from these two metrics and identify New York and London as the most central nodes in terms of degree and betweenness and, particularly, New York for the weighted degree at a global scale. However, when we restrict our analysis to just the regional scene of each community, the relative importance of each city quickly changes. The rankings for the regional weighted degree are similar to the global ones since this metric depends only on the population of each city and not on who it is connected to. The most central cities occupy the same positions except for San Diego, which slipped down three places down. On the other hand, the weighted betweenness is property that depends strongly on the network topology, a property that can be seen by the dramatic shifts we observe when considering only the local community of each city with most cities moving several positions up or down (see details in Table \ref{Betweenness} and Table S2 in Appendix). For example, San Diego went down nine places meaning that this city has a global influence due to the fact that San Diego is a communication hub between United States and Central America. Dallas went up six places, indicating that its influence is higher at the regional scale rather than in the international arena. In the same way, Madrid went down four places whereas Barcelona stayed at the same place, this means that Madrid is more influential than Barcelona at a global scale as an international bridge connecting Europe and Central and South America but not at a regional (European) scale.

\section*{DISCUSSION}

The study of competition and interactions between cities has a long history in fields such as Geography, Spatial Economics and Urbanism. This research has traditionally taken  as basis  information on finance exchanges, sharing of firm headquarters, number of passengers transported by air or tons of cargo dispatched from one city to another. One can define a network relying on these data and identify  the so-called World Cities, those with a higher level of centrality as the global economic or logistic centers. Here, we have taken a radically different approach to measure quantitatively the influence of a city in the world. Nowadays, geolocated devices generate a large quantity of real time and geolocated data permitting the characterization of people mobility. We have used Twitter data to track users and classify cities according to the mobility patterns of their visitors.
 Top cities as mobility sources or attraction points are identified as central places at a global scale for cultural and information interchanges. This definition of city influence makes possible its direct measurement instead of using indirect information such as firm headquarters or direct flights. Still, the quality of the results depends on the capacity of geolocated tweets to describe local and global mobility. Indeed, observing the World through Twitter data can lead to possible distortions, economic and  sociodemographic biases, the Twitter penetration rate may also vary from country to country leading to an under-representation of the population, for example, from Africa and from China.
The cities selected for this work are those that, on one hand, concentrate large populations and, on the other, enough number of tweets to be part of the analysis. There are biases acting against our work, as the lack of coverage in some areas of the world, and others in favor, such as the fact that younger and wealthier individuals are more likely to both travel and use Twitter.   The estimated mobility patterns are naturally partial since only refer to the selected cities. Still, as long as the users provide a significant sample of the external urban mobility, the flow network is enough for the performed analysis. Furthermore, several recent works have proven the capacity of geolocated tweets to describe human mobility comparing different data sources as information collected from cell phone records, Twitter, traffic measure techniques and surveys \cite{lenormand14,tizzoni14,lenormand14b}.  

More specifically and assuming data reliability, we consider the users' displacements after visiting each city. The urban areas are ranked according to the area covered and the radius traveled by these users as a function of time. These metrics  are inspired by the framework developed for random walks and Levy flights, which allows us to characterize the evolution of the system with well-defined mathematical tools and with a clear reference baseline in mind. Previous literature rankings usually find a hierarchy  captained by New York and London as the most central world cities. The ranks dramatically change when one has into account users' mobility. A triplet formed by Rome, Paris and Lisbon consistently appear on the top of the ranking by extension of visitor's mobility but also by their attractiveness to travelers of very diverse origin. A combination of economic activity appealing to tourism and diversity of links to other lands, in some cases product of recent history, can explain the presence of these cities on the top. These three cities are followed by others such as San Francisco that without being one of the most populated cities in the US extends it influence over the large Pacific basin or Hong Kong, Beijing and Shanghai that replicates it on the other side of the Pacific region. These cities are in some cases gates to broad hinterlands. This is relevant since our metrics have into account the diversity in the visitors' origins. 

These results rely on the full users population, discriminating only by the place of residence between locals and non locals to each city. The influence of cities measured in this way includes their impact in rural as well as in other urban areas. However, the analysis can be restricted to users residing in an urban area and to their displacements toward other cities. In this way, we obtain a weighted directed network between cities, whose links weights represent the (normalized) fluxes of users traveling from one city to another. This network provides the basis for a more traditional centrality analysis, in which we recover London and New York as the most central cities at a global scale. The match between our results and those from previous analysis brings further confidence on the quality of the flow measured from online data. The network framework permits to run clustering techniques and divide the world city network in communities or areas of influence. When the centrality is studied only within each community, we obtain a regional perspective that induces a new ranking of cities. The comparison between the global and the regional ranking provides important insights in the change of roles of cities in the hierarchies when passing from global to regional. 

Summarizing, we have introduced a new method to measure the influence of cities based on the Twitter user displacements as proxies for the mobility flows. The method, despite some possible biases due to the population using online social media, allows for a direct measurement of a city influence in the world. We proposed three types of rankings capturing different perspectives: rankings based on ``city-to-world'' and ``world-to-city'' interactions and rankings based on ``city-to-city'' interaction. It is  interesting to note that the most influential cities are very different according to the perspective and the scale (regional and global). This introduces the possibility of studying relations among cities and between cities and rural areas with unprecedented detail and scale.

\section*{ACKNOWLEDGEMENTS}

Partial financial support has been received from the Spanish Ministry of Economy (MINECO) and FEDER (EU) under projects MODASS (FIS2011-24785) and INTENSE@COSYP (FIS2012-30634),  and from the EU Commission through projects EUNOIA, LASAGNE and INSIGHT. The work of ML has been funded under the PD/004/2013 project, from the Conselleria de Educación, Cultura y Universidades of the Government of the Balearic Islands and from the European Social Fund through the Balearic Islands ESF operational program for 2013-2017. JJR from the Ram\'on y Cajal program of MINECO. BG was partially supported by the French ANR project HarMS-flu (ANR-12-MONU-0018).

\newpage
\clearpage
\newpage

\makeatletter
\renewcommand{\fnum@figure}{\small\textbf{\figurename~S\thefigure}}
\makeatother
\setcounter{figure}{0}

\makeatletter
\renewcommand{\fnum@table}{\small\textbf{\tablename~S\thetable}}
\renewcommand{\tablename}{Table}
\makeatother
\setcounter{table}{0}

\onecolumngrid

\section*{APPENDIX}

As a first characterization of the data,  we have computed the great circle distance $\Delta_r$ between successive positions of the same Twitter user living in one of the $58$ cities (Figure S1). The distribution $P(\Delta_r)$ for each city is well approximated by a power law with an average exponent value of $1.5$. These results are consistent with the exponent obtained in other studies \cite{brockmann06,gonzalez08,hawelka13}. It is interesting to note that the distributions are very similar for all the cities.

\begin{figure}[!ht]
  \begin{center}
		\includegraphics[scale=0.7]{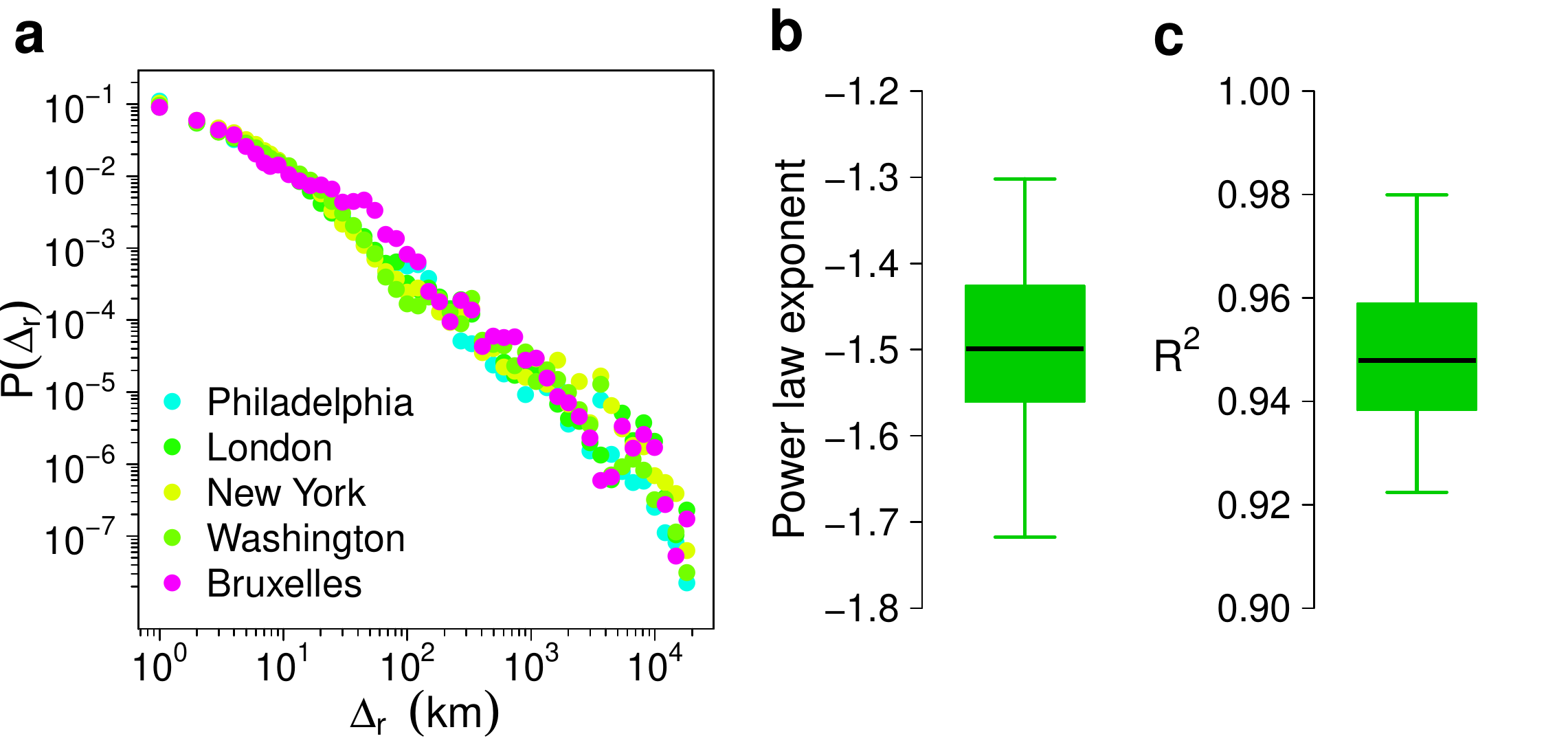}
	\end{center}
		\caption{\textbf{Probablity density function of distance travelled by the local Twitter users.} (a) Probablity density function $P(\Delta_r)$ of the distance travelled by the local Twitter users for 5 cities drawn at random among the 58 case studies. $\Delta_r$ is the great circle distance between each successive position of the local Twitter users. (b) Boxplot of the $58$ power-law exponent. (c) Boxplot of the $R^2$. The boxplot is composed of the minimum value, the lower hinge, the median, the upper hinge and the maximum value.\label{FigS1}}
\end{figure}

\begin{figure}[!ht]
  \begin{center}
		\includegraphics[scale=0.55]{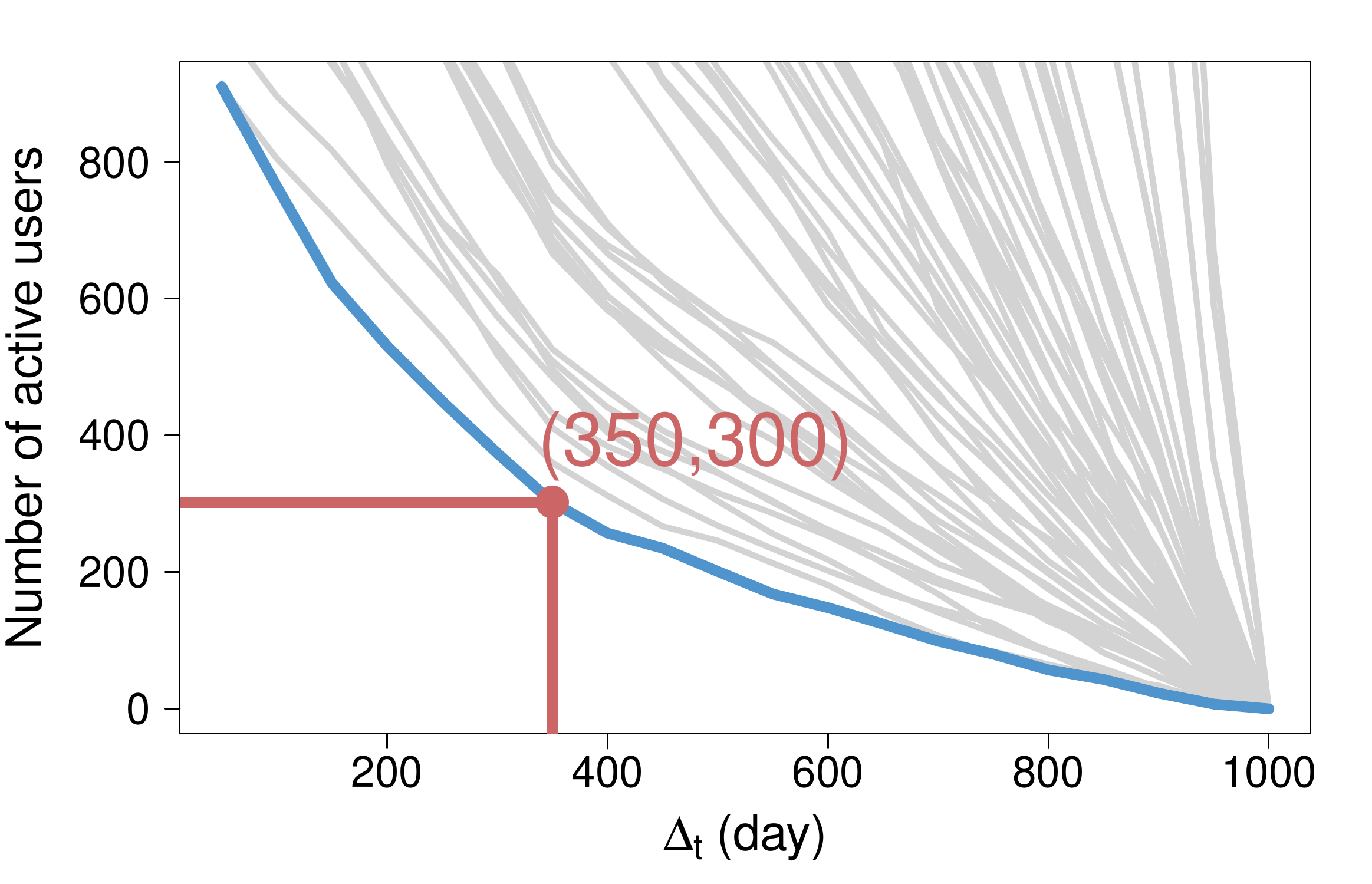}
	\end{center}
		\caption{\textbf{Minimum number of active users as a function of $\Delta_t$ (blue line).} The gray lines represent the number of active users as a function of $\Delta_t$ for the $58$ cities.\label{FigS2}}
\end{figure}

\begin{figure}
  \begin{center}
		\includegraphics[width=\linewidth]{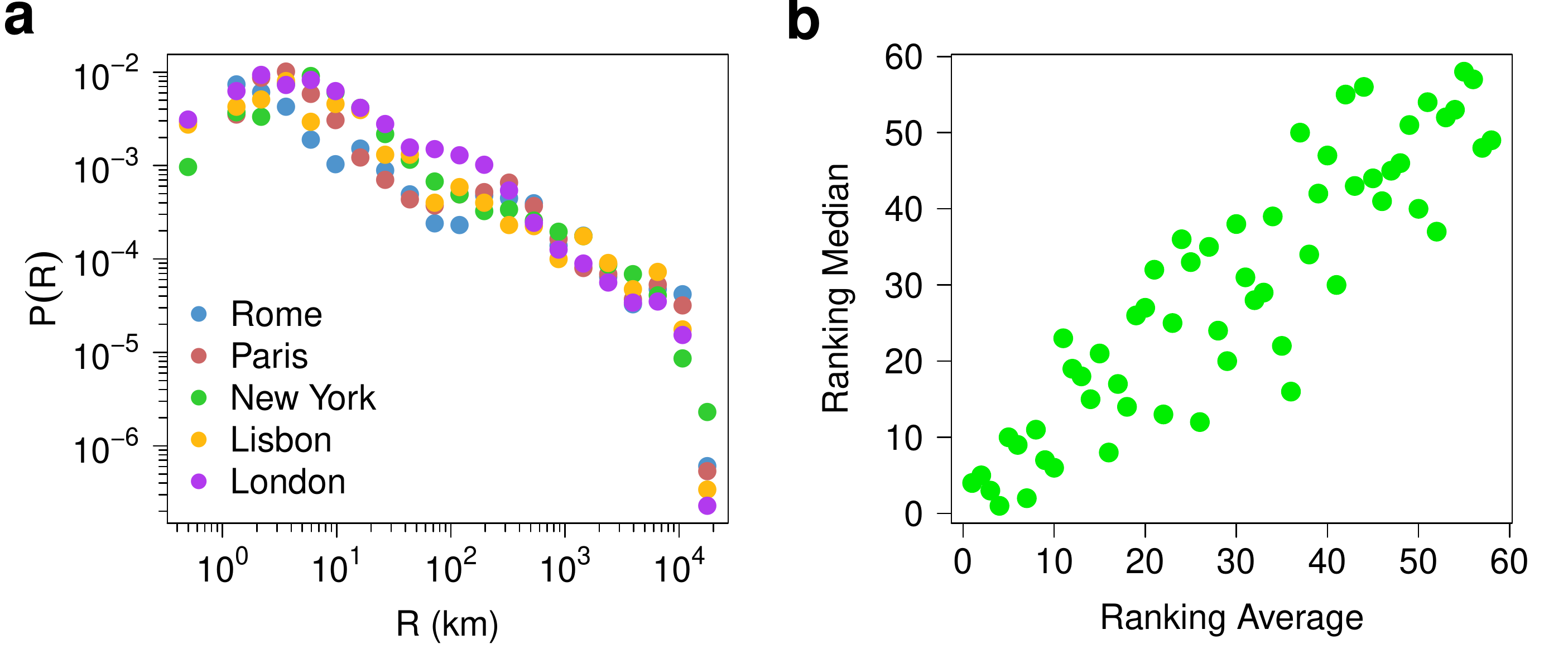}
	\end{center}
		\caption{\textbf{Radius.} (a) Probablity density function of the radius per Twitter users for 5 cities. (b) Ranking by median radius as a function of the ranking by average radius. The rankings are based on an average of the two statistics over $100$ independent extractions of a set of $u = 300$ users. \label{FigS3}}
\end{figure}

\begin{figure}
  \begin{center}
		\includegraphics[scale=0.7]{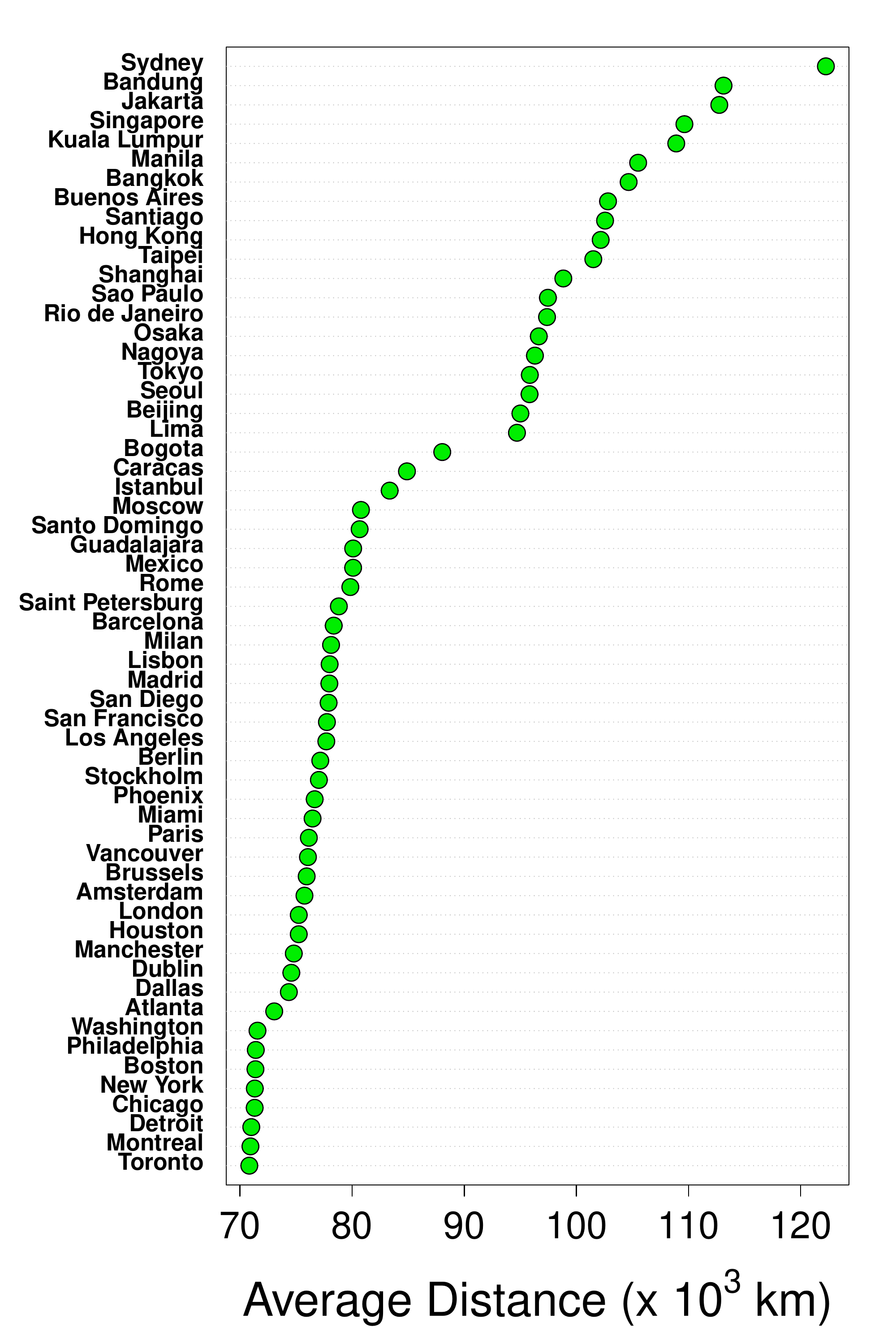}
	\end{center}
		\caption{\textbf{Ranking of the cities according the the average distance between the center of the city and all the Twitter users' place of residence (represented by the centroid of the cell of residence).}\label{FigS4}}
\end{figure}

\begin{figure}
  \begin{center}
		\includegraphics[width=\linewidth]{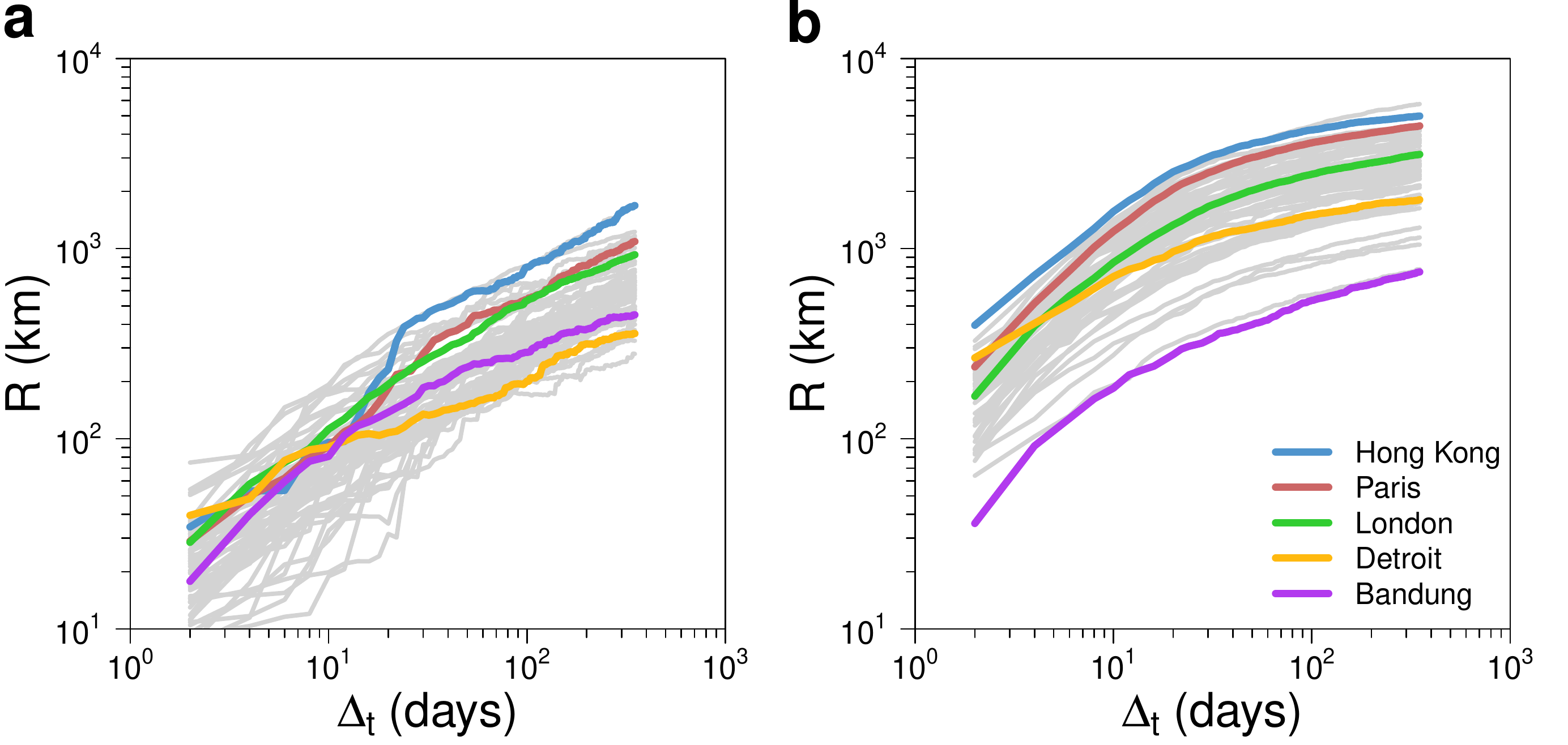}
	\end{center}
		\caption{\textbf{Evolution of the average radius for the local users (a) and for the non-local users (b).} Each curve represents the evolution of the average radius $R$ averaged over $100$ independent extractions of a set of $u = 100$ users  as a function of the number of days $\Delta_t$ since the first passage in the city. In order to show the general trend, each gray curve corresponds to a city. The evolution of the radius for several cities is highlighted, such as the top and bottom rankers or representatives of the two main detected behaviors.  \label{FigS5}}
\end{figure}

\begin{figure}
  \begin{center}
		\includegraphics[scale=0.7]{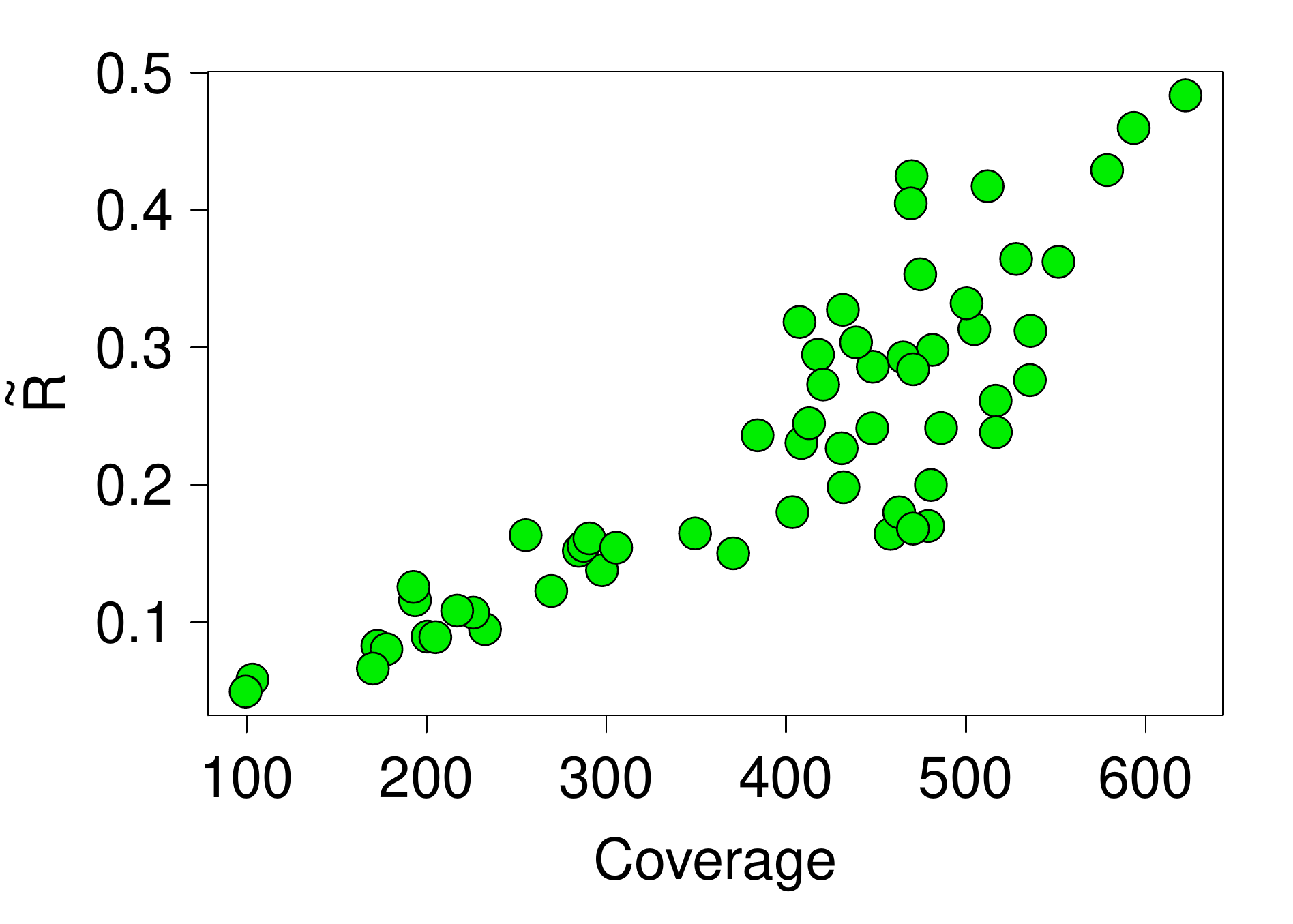}
	\end{center}
		\caption{\textbf{Coverage as a function of $\tilde{R}$ for the $58$ cities.} A certain level of correlation can be observed between both metrics. Both metrics are averaged over $100$ independent extractions of a set of $u = 300$ users.\label{FigS6}}
\end{figure}

\begin{figure}
  \begin{center}
		\includegraphics[width=\linewidth]{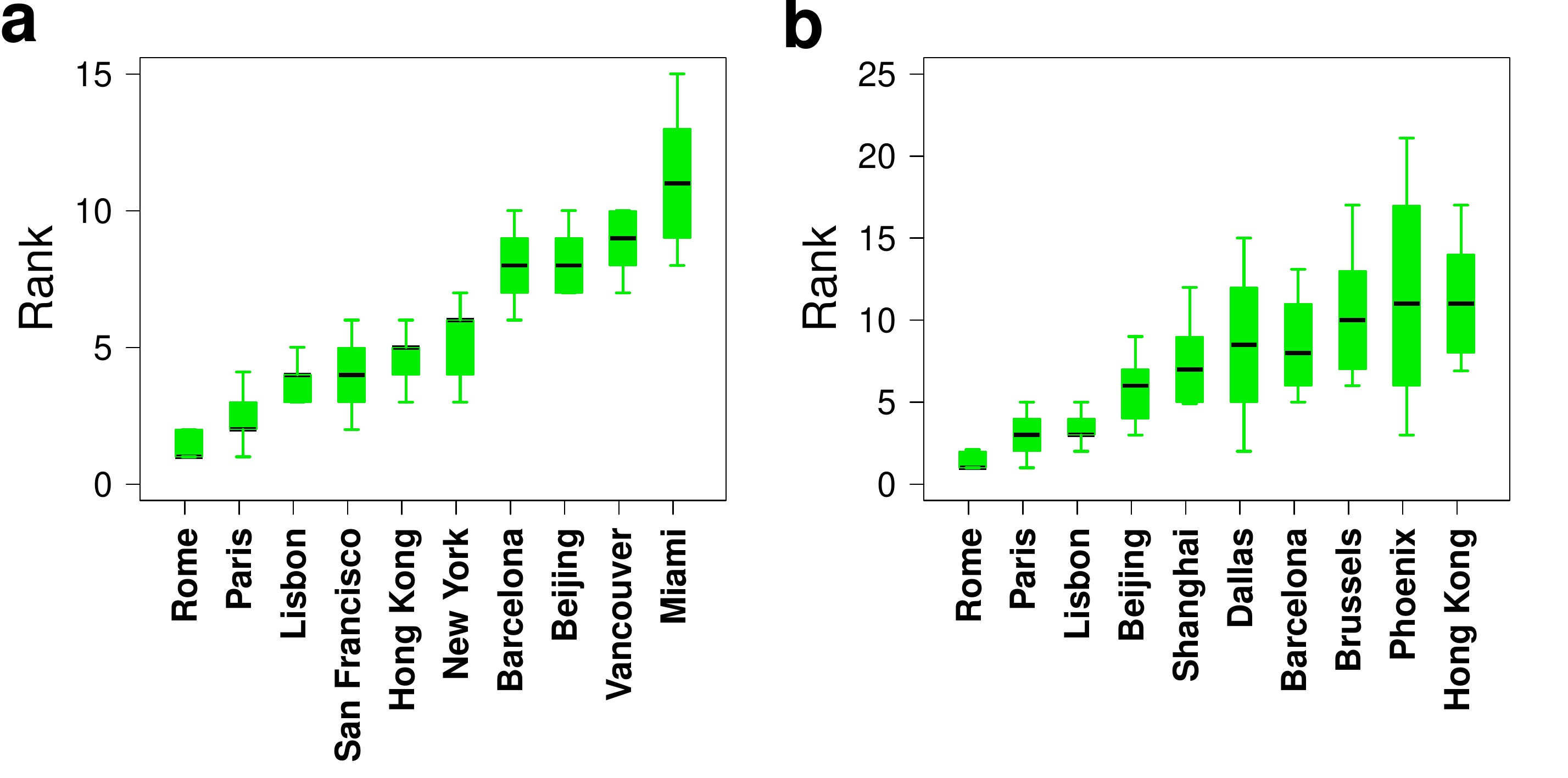}
	\end{center}
		\caption{\textbf{Variations of the rankings over 100 realizations.} (a) Ranking for the normalized average radius. (b) Ranking for the coverage. The boxplot is composed of the minimum value, the lower hinge, the median, the upper hinge and the maximum value. The rankings are averaged over $100$ independent extractions of a set of $u = 300$ users.\label{FigS7}}
\end{figure}

\begin{figure}
  \begin{center}
		\includegraphics[width=\linewidth]{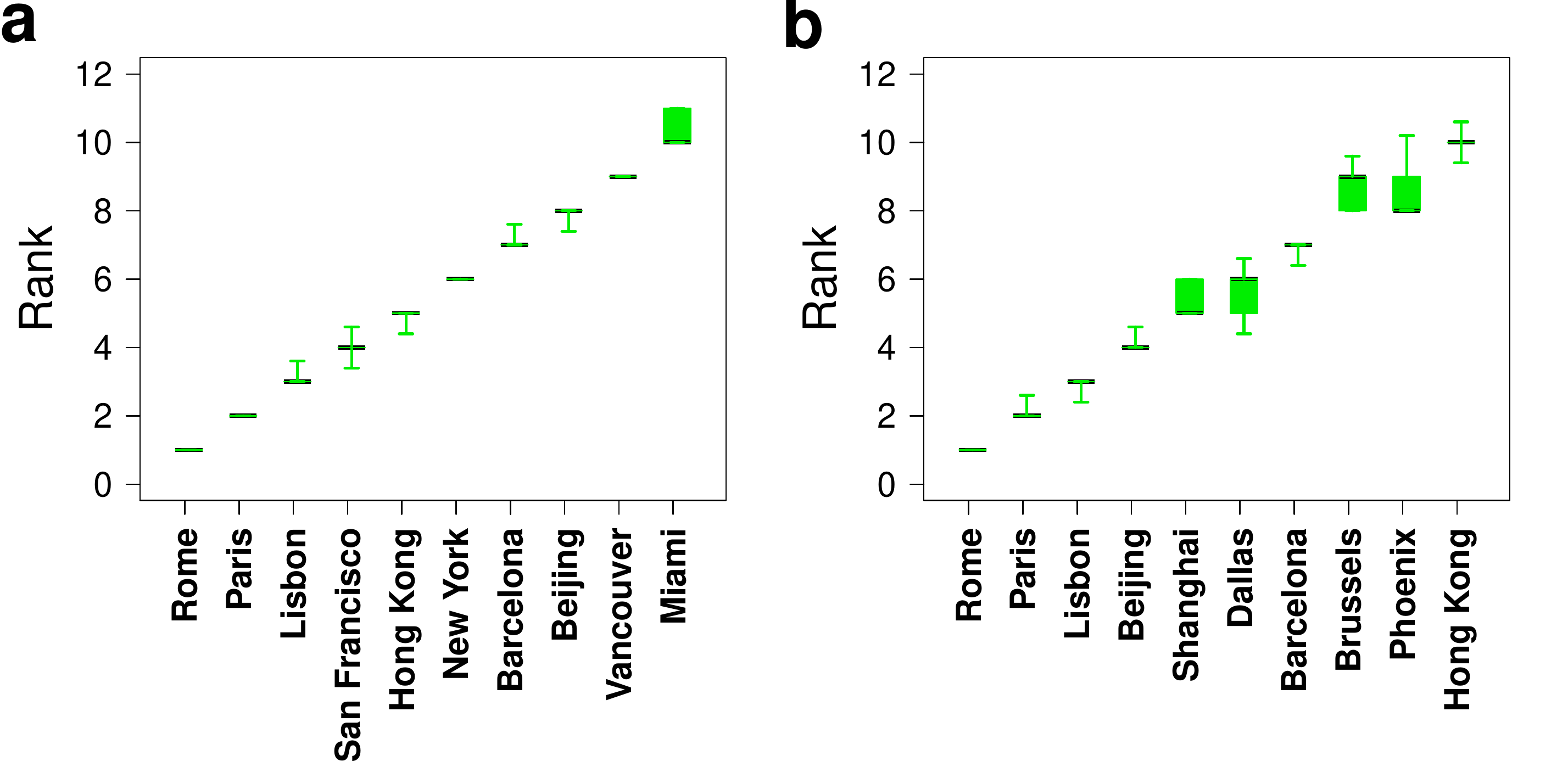}
	\end{center}
		\caption{\textbf{Variations of the rankings over 10 realizations performed on the average over 10 realizations.} (a) Ranking for the normalized average radius. (b) Ranking for the coverage. The boxplot is composed of the minimum value, the lower hinge, the median, the upper hinge and the maximum value. The rankings are averaged over $100$ independent extractions of a set of $u = 300$ users.\label{FigS8}}
\end{figure}
		
\clearpage
\newpage
\section*{Entropy index}

The natural way of taking the heterogeneity of visiting frequencies into consideration is to introduce an entropy measure. If we define the probability $p_i^t$ than an individual tweet originating from the users we are considering originated in a cell $i$, then the entropy $S$ for a given time interval $\Delta t$ is given by:

\begin{equation}
   S\left(t\right)=-\frac{\sum_{i=1}^N p_i^t\log(p_i^t)}{\log\left(\mathcal{N}\left(t\right)\right)}  
 \label{Ent}
\end{equation}

where the normalizing factor $\mathcal{N}\left(t\right)$, the number of cells with non-zero number of tweets, corresponds to the uniform case where each tweet has the same probability of being produced within each cell. With this normalization, the entropy is defined to vary just between $0$ and $1$, regardless of the number of cells and tweets we might consider in each case.

The entropy as a function of the number of visited cells is plotted in Figure \ref{FigS7}a. The entropy enhances with the number of visited cells despite the normalization, which implies that the tweets tend to distribute more uniformly for those cities with larger areas covered and therefore with a larger global projection. Besides the general trend, there are some interesting outliers such as Moscow and Saint Petersburg, with a high area covered given the size of Russia but low entropy meaning that the travels concentrate toward a few cells (likely the cities in a vast territory). On the other extreme, we find Osaka and Nagoya with a low are covered but high entropy. A possible reason is that the travels can be mostly within Japan but since the population in the country is well distributed, the trip destinations are well mixed. 

As can be seen in Figure \ref{FigS7}b, the entropy measured in the cities based only in local users is way lower than for the non-locals. This means that the locals move toward more concentrated locations, in contrast to the comparatively higher diversity of origins of the non-local visitors.

\begin{figure}[!ht]
  \begin{center}
		\includegraphics[width=\linewidth]{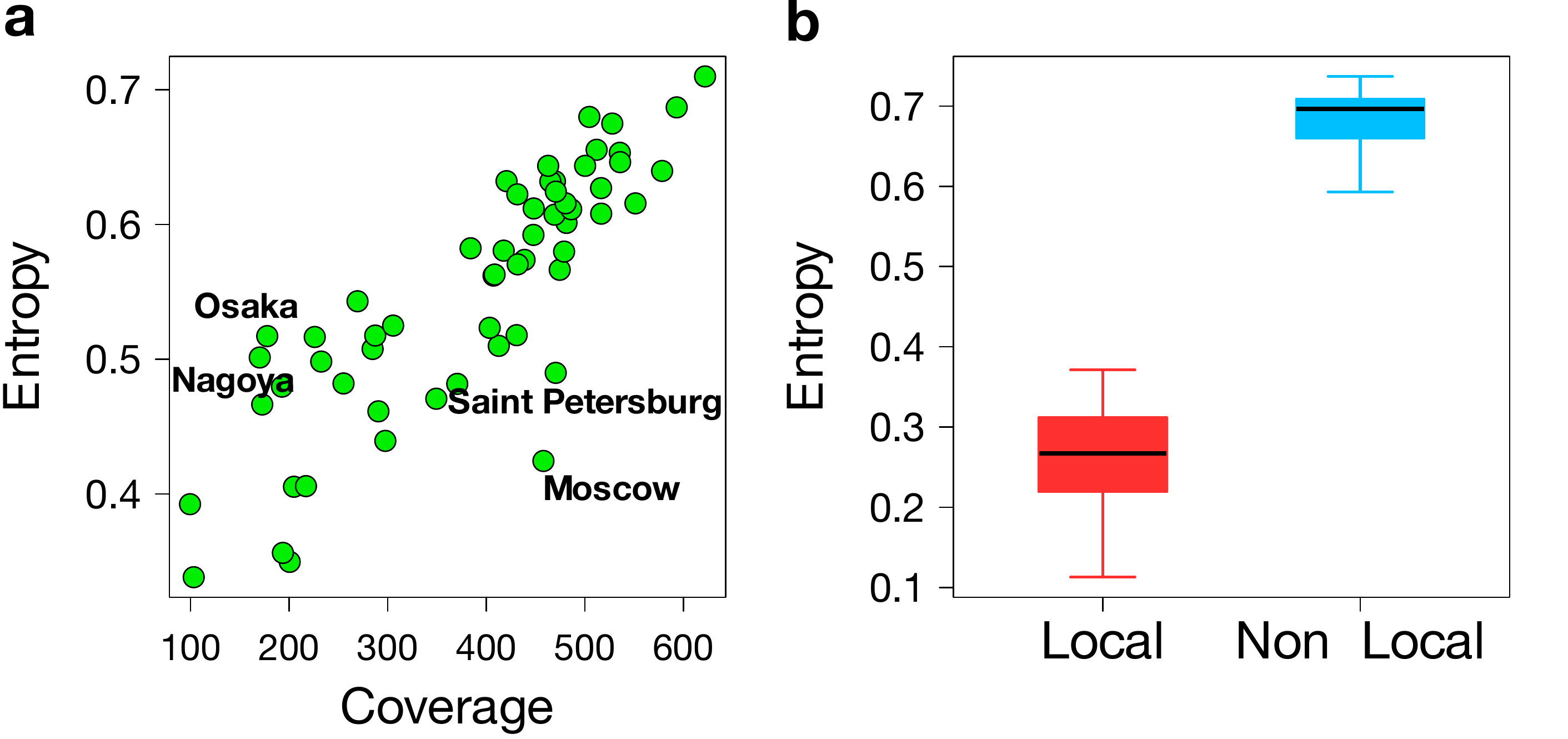}
	\end{center}
\caption{\textbf{Entropy index according to the Twitter user type.} (a) Entropy index as a function of the number of cells visited by $u=300$ Twitter users drawn at random. (b) Box plot with the entropy measured for the different cities separating the users as locals and non-locals. The number of users is $u = 100$ in this case. \label{FigS9}}
\end{figure}

\clearpage
\begin{figure}
  \begin{center}
		\includegraphics[width=\linewidth]{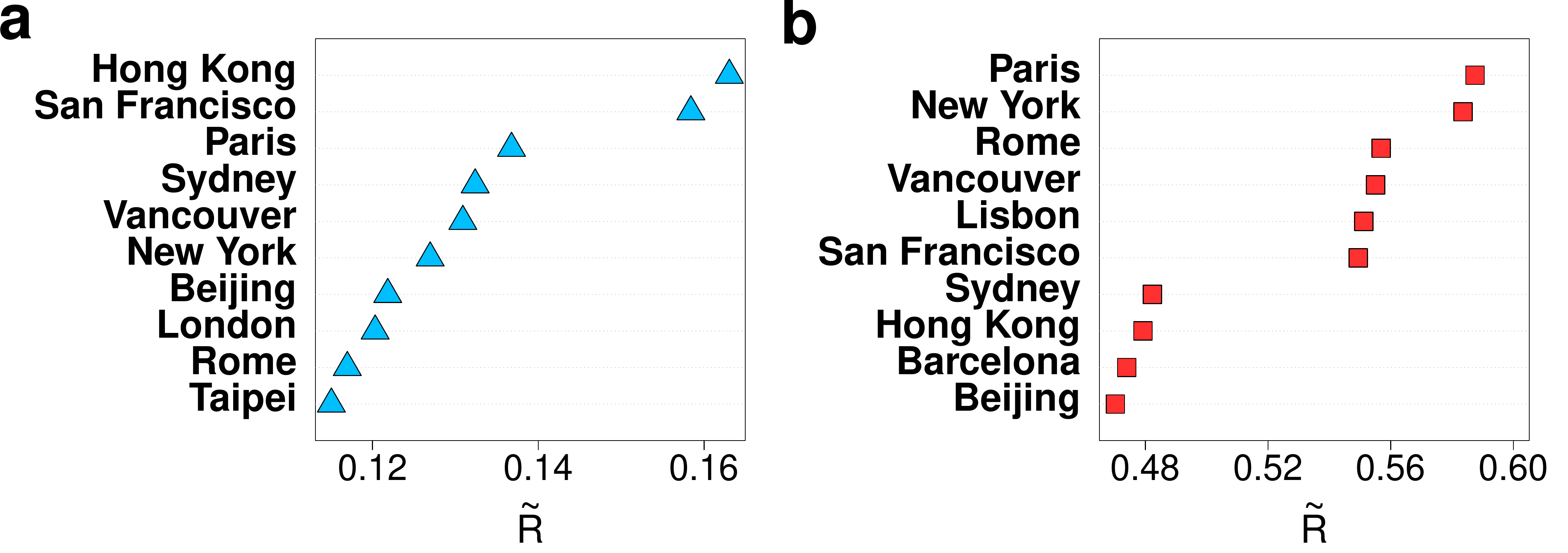}
	\end{center}
\caption{\textbf{Relation between local and non-local users.} (a) Top 10 ranking cities based only on local users according to the average radius. (b) Top 10 ranking cities based only on non-local users according to the average radius. In all the cases, the number of local and non-local users extracted is $u = 100$ for every city and all the metrics are averaged over $100$ independent extractions. \label{FigS10}}
\end{figure}

\begin{figure}
  \begin{center}
		\includegraphics[scale=0.6]{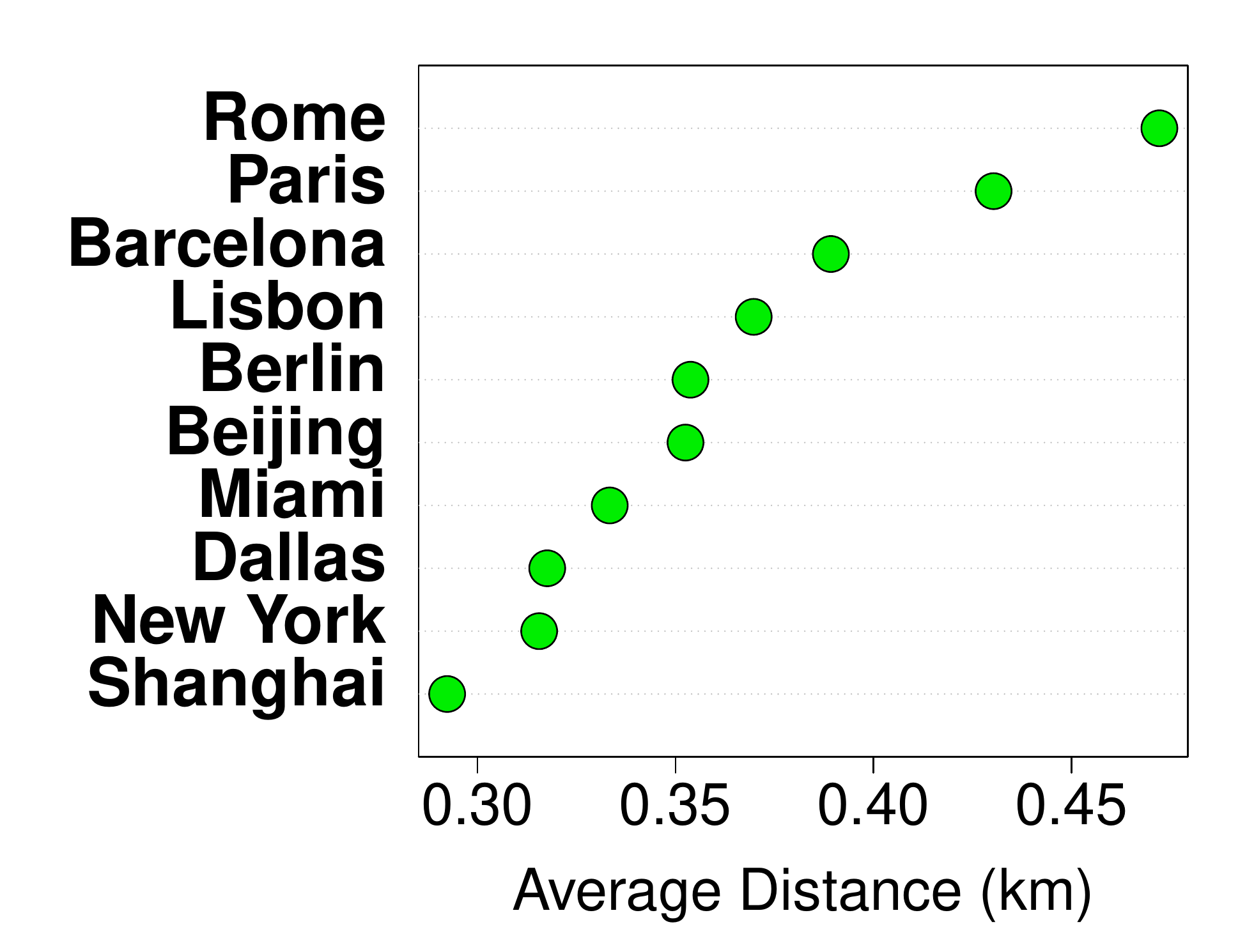}
	\end{center}
\caption{\textbf{City attractiveness.} Top 10 cities ranked by the average distance between the Twitter users' residences (represented by the centroid of the cell of residence) and the city center for $u=1000$ Twitter users drawn at random. The metric is averaged over $100$ independent extractions. \label{FigS11}}
\end{figure}
\clearpage

\newpage
\begin{table}
	\caption{\textbf{Description of the case studies}}
	\label{city}
		\begin{center}
		  \fontsize{9}{9}\selectfont
      \begin{tabular}{>{}m{3cm}>{\centering}m{2cm}>{\centering}m{2cm} m{2cm}<{\centering}}	
				\hline
				\textbf{City} & \textbf{Number of users} & \textbf{Number of Tweets} & \textbf{Number of Tweets per user}\\
				\hline
Amsterdam	&	2661	&	305363	&	114.75	\\
Atlanta	&	2863	&	296390	&	103.52	\\
Bandung	&	5620	&	405241	&	72.11	\\
Bangkok	&	2604	&	239514	&	91.98	\\
Barcelona	&	1713	&	165934	&	96.87	\\
Beijing	&	1299	&	131922	&	101.56	\\
Berlin	&	678	&	45238	&	66.72	\\
Bogota	&	2226	&	213739	&	96.02	\\
Boston	&	752	&	73561	&	97.82	\\
Brussels	&	1243	&	97688	&	78.59	\\
Buenos Aires	&	411	&	28500	&	69.34	\\
Caracas	&	3625	&	375933	&	103.71	\\
Chicago	&	2191	&	257572	&	117.56	\\
Dallas	&	1214	&	128834	&	106.12	\\
Detroit	&	13608	&	938524	&	68.97	\\
Dublin	&	704	&	78434	&	111.41	\\
Guadalajara	&	721	&	57031	&	79.10	\\
Hong Kong	&	1098	&	108203	&	98.55	\\
Houston	&	1582	&	186830	&	118.10	\\
Istanbul	&	1321	&	103117	&	78.06	\\
Jakarta	&	1919	&	196188	&	102.23	\\
Kuala Lumpur	&	509	&	42665	&	83.82	\\
Lima	&	360	&	42186	&	117.18	\\
Lisbon	&	6782	&	698998	&	103.07	\\
London	&	6392	&	580084	&	90.75	\\
Los Angeles	&	1760	&	159781	&	90.78	\\
Madrid	&	1566	&	202650	&	129.41	\\
Manchester	&	1792	&	163090	&	91.01	\\
Manila	&	4118	&	293015	&	71.15	\\
Mexico	&	2534	&	247486	&	97.67	\\
Miami	&	688	&	84544	&	122.88	\\
Milan	&	666	&	61175	&	91.85	\\
Montreal	&	1239	&	133461	&	107.72	\\
Moscow	&	2334	&	263132	&	112.74	\\
Nagoya	&	9668	&	892442	&	92.31	\\
New York	&	4044	&	398769	&	98.61	\\
Osaka	&	2567	&	247449	&	96.40	\\
Paris	&	432	&	43301	&	100.23	\\
Philadelphia	&	2206	&	247159	&	112.04	\\
Phoenix	&	1380	&	150468	&	109.03	\\
Rio de Janeiro	&	3292	&	352777	&	107.16	\\
Rome	&	824	&	88402	&	107.28	\\
Saint Petersburg	&	497	&	51601	&	103.82	\\
San Diego	&	1810	&	182035	&	100.57	\\
San Francisco	&	4628	&	419032	&	90.54	\\
Santiago	&	2471	&	250639	&	101.43	\\
Santo Domingo	&	302	&	20245	&	67.04	\\
Sao Paulo	&	6479	&	653909	&	100.93	\\
Seoul	&	1898	&	152666	&	80.44	\\
Shanghai	&	526	&	49282	&	93.69	\\
Singapore	&	3501	&	288267	&	82.34	\\
Stockholm	&	745	&	106366	&	142.77	\\
Sydney	&	1176	&	121426	&	103.25	\\
Taipei	&	485	&	40259	&	83.01	\\
Tokyo	&	10333	&	844602	&	81.74	\\
Toronto	&	1476	&	135914	&	92.08	\\
Vancouver	&	796	&	70018	&	87.96	\\
Washington	&	3755	&	421374	&	112.22	\\
				\hline
	  	\end{tabular}
	  \end{center}
\end{table}

\newpage
\begin{table}
	\caption{\textbf{Comparison of the regional and the global betweenness rankings.}}
	\label{Betweenness}
		\begin{center}
		  \fontsize{9}{9}\selectfont
      \begin{tabular}{|>{}m{3cm}|>{}m{4cm}| m{4cm}<{}|}	
				\hline
				\textbf{Community} & \textbf{Global Ranking} & \textbf{Regional Ranking}\\
				\hline
North America	&	1. New York (1)	&	1. New York	\\
	&	2. Miami (6)	&	2. Los Angeles	\\
	&	3. San Francisco (8)	&	3. Chicago	\\
	&	4. Los Angeles (9)	&	4. Toronto	\\
	&	5. Chicago (18)	&	5. Detroit	\\
	&	6. Toronto (19)	&	6. Miami	\\
	&	7. San Diego (23)	&	7. Dallas	\\
	&	8. Detroit (25)	&	8. San Francisco	\\
	&	9. Montreal (26)	&	9. Washington	\\
	&	10. Atlanta (27)	&	10. Atlanta	\\
	&	11. Washington (29)	&	11. Phoenix	\\
	&	12. Vancouver (35)	&	12. Vancouver	\\
	&	13. Dallas (36)	&	13. Montreal	\\
	&	14. Phoenix (46)	&	14. Boston	\\
	&	15. Boston (47)	&	15. Houston	\\
	&	16. Houston (48)	&	16. San Diego	\\
	&	17. Philadelphia (50)	&	17. Philadelphia	\\
	&	18. Santo Domingo (58)	&	18. Santo Domingo	\\
	\hline				
Europe	&	1. London (2)	&	1. London	\\
	&	2. Paris (3)	&	2. Paris	\\
	&	3. Madrid (10) &	3. Moscow	\\
	&	4. Barcelona (11)	&	4. Barcelona	\\
	&	5. Moscow (16) &	5. Berlin	\\
	&	6. Berlin (20) &	6. Rome	\\
	&	7. Rome (21) &	7. Madrid	\\
	&	8. Amsterdam (24) &	8. Lisbon	\\
	&	9. Lisbon (38) &	9. Amsterdam	\\
	&	10. Milan (40) &	10. Saint Petersburg	\\
	&	11. Brussels (41) &	11. Dublin	\\
	&	12. Istanbul (42) &	12. Istanbul	\\
	&	13. Saint Petersburg (45)	&	13. Manchester	\\
	&	14. Dublin (49) &	14. Brussels	\\
	&	15. Manchester (51) &	15. Milan	\\
	&	16. Stockholm	(57) &	16. Stockholm	\\
	\hline				
Asia	&	1. Singapore (5) &	1. Singapore	\\
	&	2. Hong Kong (7) &	2. Hong Kong	\\
	&	3. Taipei (13) &	3. Jakarta	\\
	&	4. Jakarta (15)	&	4. Bangkok	\\
	&	5. Kuala Lumpur (22)	&	5. Shanghai	\\
	&	6. Seoul (30)	&	6. Taipei	\\
	&	7. Bangkok (31)	&	7. Sydney	\\
	&	8. Shanghai (32) &	8. Kuala Lumpur	\\
	&	9. Beijing (33)	&	9. Seoul	\\
	&	10. Sydney (34)	&	10. Manila	\\
	&	11. Manila (43)	&	11. Bandung	\\
	&	12. Bandung	(56) &	12. Beijing	\\
	\hline				
South America	&	1. Buenos Aires (12)	&	1. Buenos Aires	\\
	&	2. Sao Paulo (14) &	2. Sao Paulo	\\
	&	3. Bogota	(28) &	3. Bogota	\\
	&	4. Santiago	(37) &	4. Rio de Janeiro	\\
	&	5. Rio de Janeiro (39)	&	5. Santiago	\\
	&	6. Lima (44)	&	6. Caracas	\\
	&	7. Caracas (55)	&	7. Lima	\\
	\hline				
Japan	&	1. Tokyo (4) &	1. Tokyo	\\
	&	2. Osaka (53)	&	2. Osaka	\\
	&	3. Nagoya (54) &	3. Nagoya	\\
	\hline				
Mexico	&	1. Mexico (17) &	1. Guadalajara	\\
	&	2. Guadalajara (52) &	2. Mexico	\\
	\hline				
			
	  	\end{tabular}
	  \end{center}
\end{table}

\end{document}